# All Electrical Near-Zero Field Magnetoresistance Magnetometry up to 500°C Using SiC Devices

F. Sgrignuoli *[1], I. Viti [1], Z.G. Yu [1] , E. Allridge[2], P. Lenahan [2], S. Goswami [3], R. Ghandi [3], M. Aghayan [3], and D.M.Shaddock[3]

[1] QuantCAD LLC, 5235 S. Harper Ct., Chicago IL, 60615, USA

[2] The Pennsylvanian State University, University Park, PA, 16802, USA

[3] GE Aerospace Research Center, 1 Research Cir, Niskayuna, NY, 12309, USA

* fabrizio.sgrignuoli@quantcad.com

**Abstract.**

Silicon Carbide (SiC) is renowned for its exceptional thermal stability, making it a crucial material for high-temperature power devices in extreme environments. While optically detected magnetic resonance (ODMR) in SiC has been widely studied for magnetometry, it requires complex setups involving optical and microwave sources. Similarly, electrically detected magnetic resonance (EDMR) in SiC, which relies on an electrical readout of spin resonance, has also been explored for magnetometry. However, both techniques require microwave excitation, which limits their scalability. In contrast, SiC's spin-dependent recombination (SDR) currents enable a purely electrical approach to magnetometry through the near-zero field magnetoresistance (NZFMR) effect, where the device resistance changes in response to small magnetic fields. Despite its potential, NZFMR remains underexplored for high-temperature applications. In this work, we demonstrate the use of NZFMR in SiC diodes for high-temperature relative magnetometry and achieve sensitive detection of weak magnetic fields at temperatures up to 500°C. Our technology provides a simple and cost-effective alternative to other magnetometry architectures, eliminating the need for a microwave source or complex setup. The NZFMR signal is modulated by an external magnetic field, which alters the singlet-triplet pair ratio controlled by hyperfine interactions between nuclear and electron/hole spins, as well as dipole-dipole/exchange interactions between electron and hole spins, providing a novel mechanism for relative magnetometry sensing at elevated temperatures. A critical advantage of our approach is the sensor head's low power consumption, which is less than 0.5 W at 500°C for magnetic fields below 5 Gauss. This approach provides a sensitive, reliable, and scalable solution with promising applications in space exploration, automotive systems, and industrial sectors, where high performance in extreme conditions is essential.

## Introduction

Due to its unique thermal, mechanical, and electronic properties, SiC has emerged as a crucial material in high-power electronics and devices designed for operation in harsh environments. Its wide bandgap (approximately 3.2 eV for 4H-SiC) allows SiC to sustain high electric fields, high voltages, efficient heat dissipation, and operate at temperatures far exceeding those of silicon-based devices [1]. These attributes make this material indispensable in applications requiring durability and efficiency under extreme conditions, such as power electronics, automotive systems, smart grids, and aerospace [2-4], to cite a few. Additionally, SiC hosts defects that exhibit optically and electrically active spin states, making it a versatile material for quantum sensing and spintronics [5-11]. Unlike nitrogen-vacancy (NV) centers in diamonds, SiC defects are compatible with modern semiconductor fabrication techniques, offering a scalable platform for quantum technologies [5, 11].

A concrete example of SiC's pivotal role lies in small-size, weight, and power consumption (SWaP) magnetometry for high-temperature environments, crucial for applications ranging from engine monitoring and chip quality control to space exploration [12-14]. Conventional magnetometers, such as fluxgates, face significant limitations in size, power consumption, and susceptibility to thermal noise [14]. Fluxgate magnetometers also require frequent calibration and are prone to performance degradation due to noise interference [14]. ODMR techniques have introduced high sensitivity for magnetic field detection at smaller scales [15]. However, ODMR systems, such as optically pumped magnetometry (OPM), rely on microwave excitation and optical detection, complicating their integration into SWaP devices [14]. Similarly, EDMR, which monitors spin-dependent phenomena by detecting changes in electrical current, improves signal readout but still requires microwave fields, limiting its practicality in miniaturized devices [16]. These techniques also struggle to function in high-temperature environments, with fluxgates degrading above 150-200°C [17], OPMs suffering from instability in atomic vapor cells [18], and EDMR experiencing reduced spin coherence and increased thermal noise [19].

Near-zero-field magnetoresistance (NZFMR) is emerging as a powerful defect metrology alternative [20-23], overcoming many of the limitations of traditional magnetometry [14]. NZFMR does not require microwave sources or large magnetic fields, significantly reducing system complexity, power consumption, and cost [24]. By avoiding RF fields, NZFMR enables spectroscopy below metallization layers to detect point defects in fully processed semiconductor devices, making it highly valuable for integrated circuit reliability [23]. Moreover, NZFMR allows for self-calibrating magnetometry [14]. Similar to the EDMR, NZFMR involves spin-dependent changes in current, but instead of relying on resonance at higher fields, the change is centered at near-zero magnetic fields. Low-field hyperfine mixing and electron-electron dipolar interactions alter the singlet-to-triplet pairing ratio in an NZFMR response, enabling the detection of changes in the recombination current with external magnetic field variations [24-26]. Additionally, the magnetic isotopes of host and dopant atoms involved in hyperfine interactions act as natural, stable markers, allowing self-calibration over time and temperature [14]. This stability is rooted in the stable energy levels of SiC's defects, which remain unaffected by temperature fluctuations, ensuring consistent performance across various conditions [6].

By utilizing NZFMR to probe the spin properties of active defects in standard SiC p-n junction devices, we have demonstrated an all-electrical SWaP relative magnetometry capable of operating at temperatures up to 500°C without isotopic purification or design enhancements. The proposed technology represents a significant advancement, overcoming the limitations of existing magnetometry techniques and opening new avenues for metrology and space exploration. It provides a reliable, scalable, cost-effective solution for extreme environments without sacrificing sensitivity, setting a new standard for high-performance magnetic field sensing across the most challenging conditions.

## Methods

The core of our method relies on the near-zero-field spin-dependent recombination (SDR) phenomenon, which enables the electrical readout of a current that encodes the magnetic field in which the sensor is immersed. The intrinsic defects of the SiC device dictate the response of the SDR current. When a semiconductor junction device is biased to yield a pronounced recombination current, carriers couple with deep-level (spin-dependent) defect electrons, as schematized in Fig.1. These spin states can be singlet or triplet states. Due to the Pauli exclusion principle, electron-hole recombination is allowed only in a singlet spin configuration, creating a bottleneck when the pair is in a triplet state. At near-zero magnetic fields, this bottleneck is alleviated by low hyperfine interactions [25, 26], which flip the defect electron's spin triplet into singlet states. This process remains active despite the thermal energy being orders of magnitude larger than the single-triplet mixing energy.

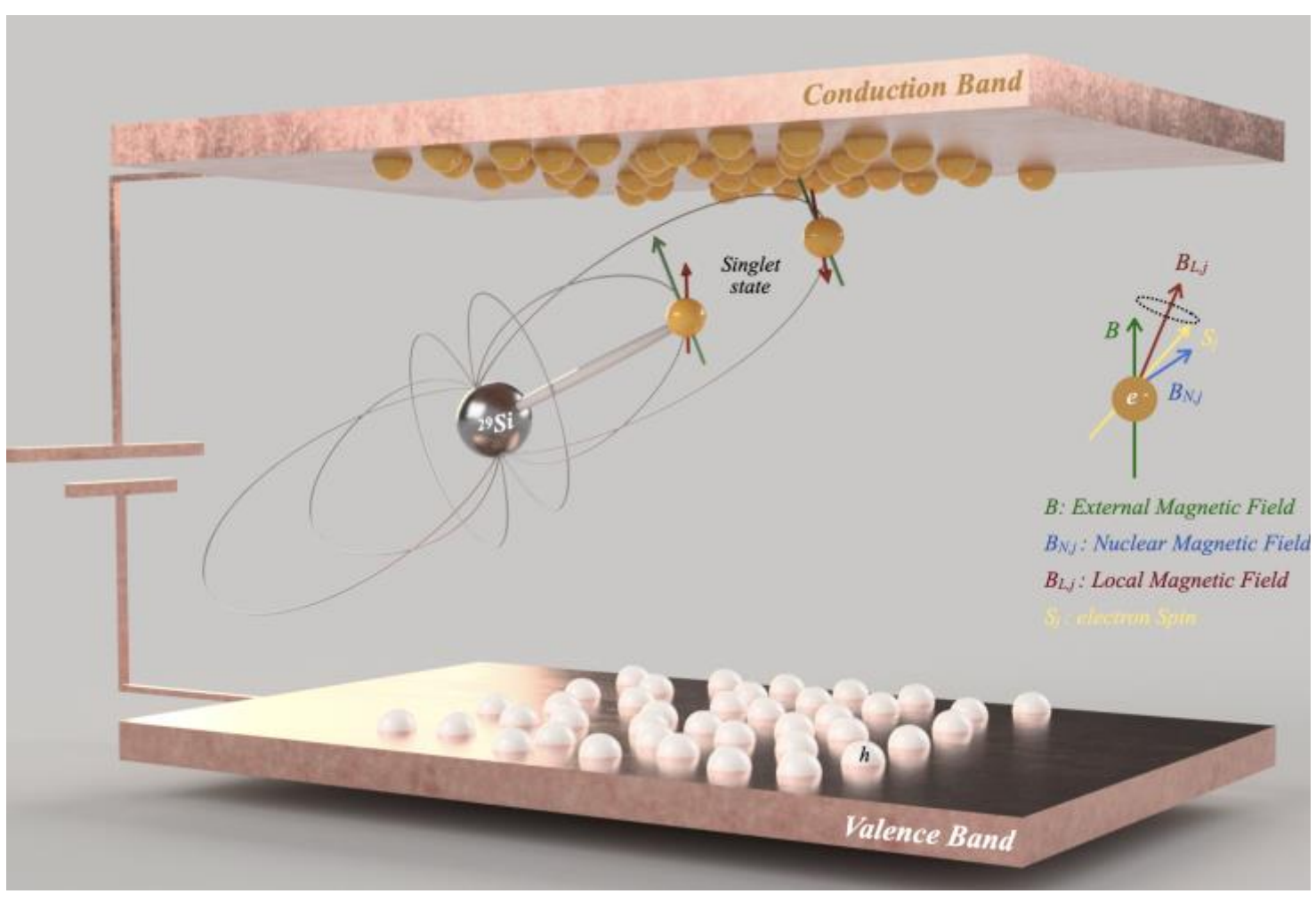

*Figure 1: Illustration of the SDR mechanism, the primary responsible for the NZFMR effect. A conduction electron (gold sphere) and a trapped electron (gold sphere bound to a $^{29}Si$ isotope, represented by a grey sphere) couple to form an intermediate spin state. The singlet-triplet pairs form due to a nuclear magnetic field $B_{N,j}$ and a small external magnetic field B. The local magnetic field sensed by each electron $S_j$ is then the vector sum of these two fields.*

Figure 2 shows our custom experimental setup for detecting DC magnetic fields via a conventional SiC diode's response using lock-in amplification. As schematized in Figure 2a, this approach relies on standard equipment and traditional SiC devices rather than isotopically purified materials, offering a cost-effective solution. The SiC device utilized in this study was not specifically designed for magnetometry. It is a lateral $p^+/i/n^+$ diode intended for high-temperature operation, featuring an ion-implanted p-well within a low-doped n-epitaxial region on a mid-cut 4H-SiC conductive $N^+$ substrate. The structure is encapsulated by vapor-deposited $SiO_2$ and $Si_3N_4$ films. The die was attached to a 500 °C–compatible alumina substrate using gold-nanoparticle sintered pastes, with Au pads and routing lines directly written onto the substrate [27–29]. This packaged diode is stacked between two sets of coils [30], functioning respectively as modulation and nulling coils. Even though the nulling coils can cancel out unwanted background magnetic fields in real-world applications, they are utilized to generate the quasi-DC magnetic field sensed in this work. A signal generator (e.g., Siglent SDG2042X) is configured to ramp the voltage applied to the nulling coil amplifier from a negative to a positive value over 60 seconds. This controlled voltage sweep alters the background magnetic field experienced by the SiC diode. Repeating this 60-second sweep multiple times, we can acquire multiple datasets under the same conditions and then average them to enhance signal quality and reduce noise. The nulling coil amplifier is assembled entirely from commercially available components—an OPA541-based, high-voltage, high-current audio amplifier board—providing a budget-friendly alternative to specialized lab instrumentation. Similarly, a TPA3116D2-based XH-M543 dual-channel stereo audio amplifier board drives the modulation coils. Both amplifiers are powered by a programmable DC power source (e.g., Rigol DP932E). The modulation coils generate alternating magnetic fields driven by an AC signal, which interact with the sample to modulate its magnetization. The modulation frequency is carefully chosen to enhance signal sensitivity significantly when suppressing low-frequency noise. Besides their different functionalities, these coils have a 28 AWG cross-sectional area and a field efficiency of 40 Gauss amps [30]. They were connected with ceramic-coated wire, secured with through-hole connectors, and further enhanced with conductive paste to reduce resistance to create the stack shown in Fig. 2b. This assembly is housed within a high-temperature oven (e.g., TableTop furnace company), showing full operation even after 60 minutes at 500°C with a current of 2 amps, demonstrating long-operation capabilities.

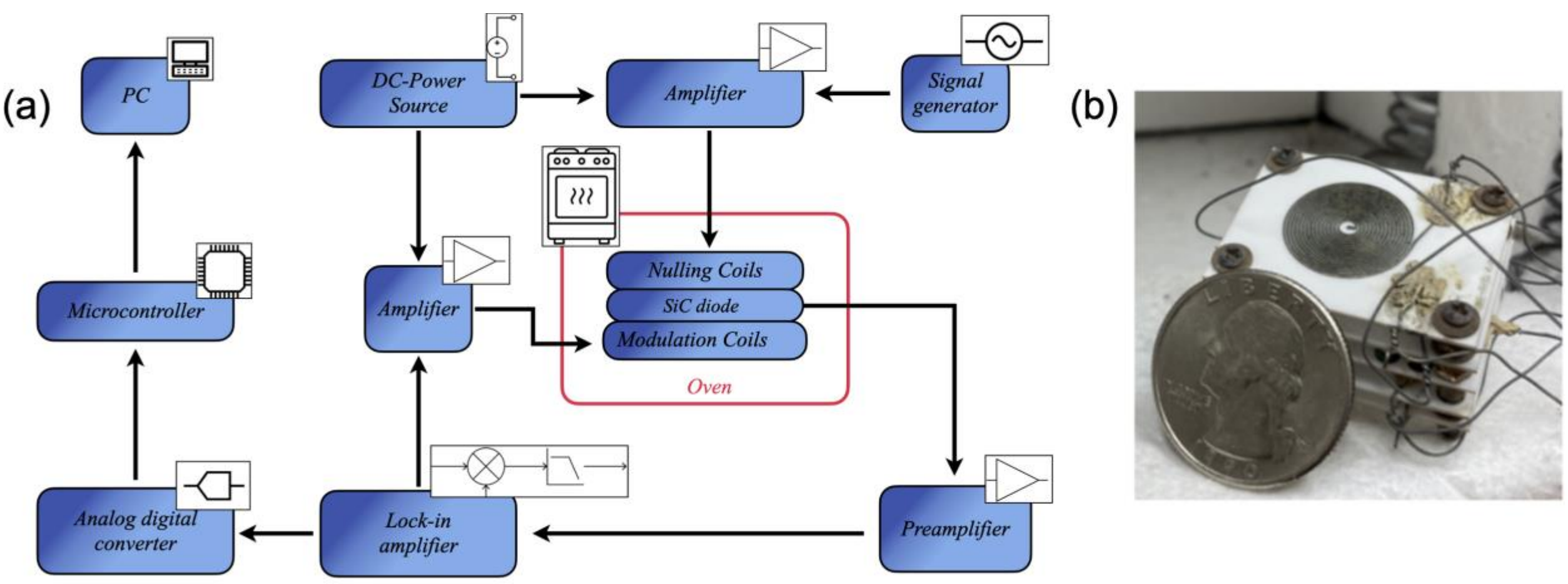

Figure 2: (a) Diagram of the high-temperature magnetometry setup. (b) A SiC diode mounted on a printed circuit board chip stack with modulation and nulling coils, specifically designed to sustain high-temperature operations, formed the assembly, which was placed in the oven. The proposed technology is approximately the size of a quarter, demonstrating a significant reduction in form factor.

To handle the weak signals produced by the SiC diode, the setup integrates a preamplifier, a lock-in amplifier (e.g., SR850 DSP model from Stanford Instruments), and different control components, as illustrated in Figure 2(a). The preamplifier strengthens the weak SDR signals, while the lock-in amplifier isolates and enhances the signal of interest by locking onto the modulation frequency. This lock-in technique improves the signal-to-noise ratio, making extracting meaningful data from the noisy environment easier. The lock-in amplifier is tuned to the modulation frequency, allowing selective detection of the resonant responses induced by the alternating magnetic field. The system also includes an analog-to-digital converter (ADC) handled by an Arduino–compatible microcontroller (e.g., ELEGOO Uno R3 board), which features a built-in 10-bit ADC, which reads the amplified analog signals and sends the digitized data to a PC interface for further analysis. In summary, the combination of high-temperature materials, advanced modulation techniques, and sensitive detection (mostly made of commercially available components assembled in-house) makes the system suitable for robust DC magnetometry in extreme conditions. Significantly, our setup is not limited to DC detection. By adequately changing the electronics and the features of our lock-in amplifier to detect signal characterized by frequencies of the order of kHz, we can also extend our method to detect AC magnetic fields.

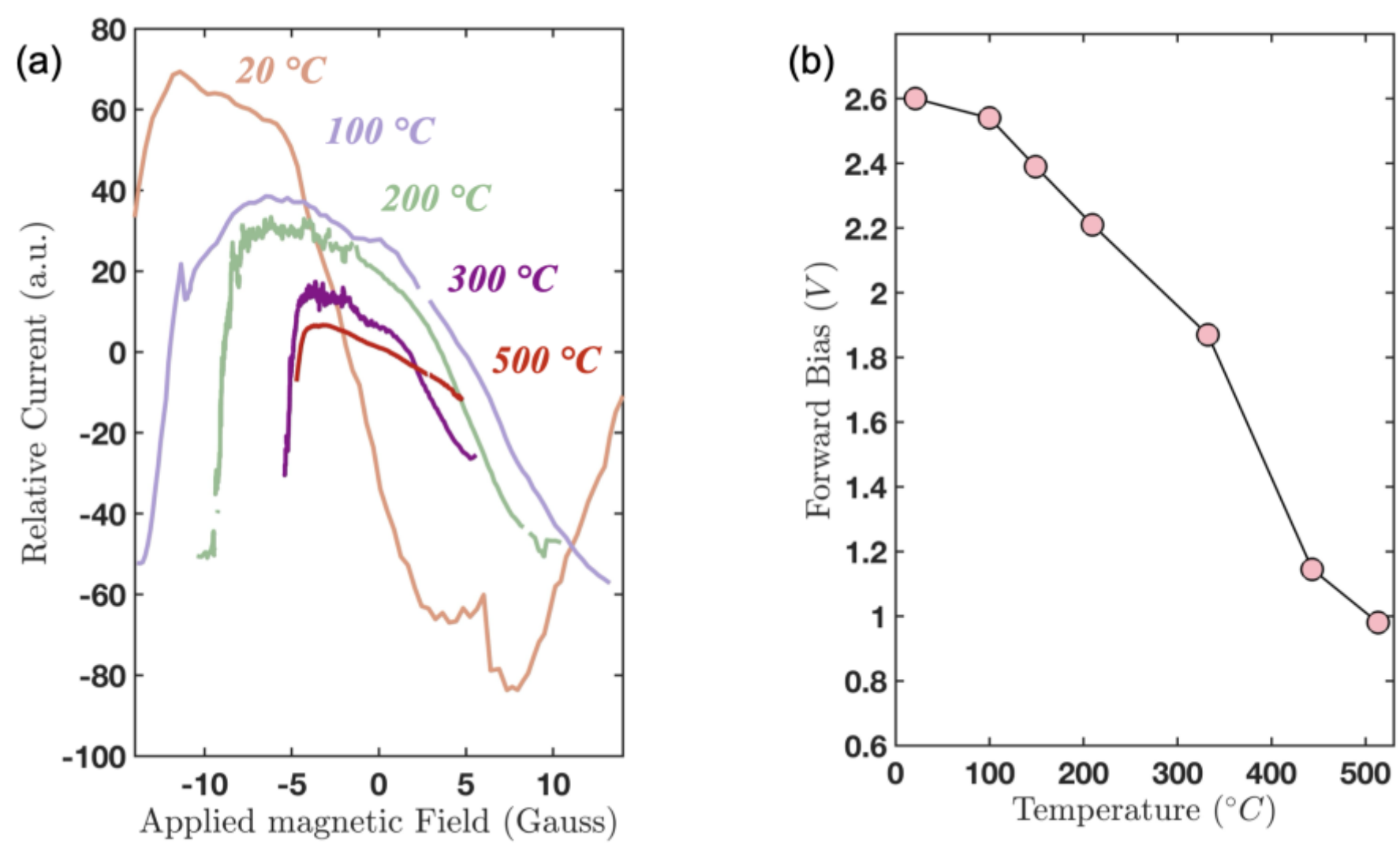

Figure 3: (a) Relative current as a function of the applied magnetic field at different temperatures. 500°C data are multiplied by a factor of 2 to increase visibility. (b) Forward bias voltage as a function of temperature. These data are obtained by fixing the diode current to 200 nA for all the different temperatures.

## Results and Discussions

Figure 3 reports averaged NZFMR signals (panel a) and forward bias (panel b )measurements for different temperatures. As the temperature increases, the NZFMR signals shift, and the relative current exhibits a gradual change across the magnetic field range (see panel (a)). This shift stems from the local change in the magnetic field background sensed by the SiC diode at higher temperatures. Because these background variations remain unquantified (and lie beyond the scope of this study, which focuses on demonstrating the reliability of relative magnetometry at high temperatures), we refer to our results as “relative.” The data of panel (a) are averaged over 20 measurements at each temperature. Without this averaging, the temperature-induced drift would mask the NZFMR signal, making it challenging to detect magnetic field-related changes. Although the NZFMR signal weakens at higher temperatures (see Fig. 3a), the correlation between the magnetic field and current remains evident, confirming the system’s ability to detect NZFMR under extreme thermal conditions. The SiC device becomes more conductive, increasing the temperatures. Indeed, the bias voltage is reduced to maintain a constant current of 200nA on a SiC diode, which helps minimize thermal effects. Also, power consumption increases slightly at 500°C due to an increase in coil resistivity from 0.5Ω at 20°C to 1.2Ω at 500 °C. Still, it remains low: lower than 0.5 W for fields less than 5 Gauss. Therefore, the proposed technology operates at low power, reflecting the vision of a robust SWaP all-electrical magnetometer on-chip to hostile environments.

Figure 4 demonstrates how an applied offset magnetic field affects NZFMR measurements and device sensitivity. Panel (a) shows NZFMR signals for three offset fields: +1.5 Gauss (blue), 0 Gauss (green), and −1.5 Gauss (red). Each trace exhibits a systematic vertical shift in relative current as the offset field varies. The red and blue curves intersect a horizontal reference line—originating from the zero-offset (green) curve—at approximately −1.25 G and +1.5 G, respectively. These values match the applied offset fields within a precision of ±0.25 G. This shift indicates a clear correlation between changes in the magnetic environment and the relative current response, demonstrating NZFMR magnetometry at high temperatures. Panel (b) presents the device's sensitivity as a function of bandwidth for various temperatures, estimated using the following relation:

$$\frac{\delta B}{\sqrt{\Delta f}} = 2\sigma\sqrt{\pi q}\,\frac{\sqrt{I_0}}{\Delta I} \qquad (1)$$

where $q$ is the electronic charge, $I_0$ is the DC current responsible for the flicker noise, $\Delta f$ is the measurement bandwidth, ΔI is the current change, and $\sigma$ is the signal line width [14]. As the bandwidth increases, the system's sensitivity decreases because more noise is introduced over a broader range of frequencies, which lowers the signal-to-noise ratio and makes it harder to detect small changes in the magnetic field. Simultaneously, panel (b) also shows a temperature dependence, with lower temperatures yielding better sensitivity. This behavior is due to increased thermal noise, raising the temperature. This reduction in sensitivity at elevated temperatures is primarily driven by two factors: the broadening of the NZFMR signal (i.e., σ) and a decrease in the current response (i.e., ΔI) to magnetic fields. Higher temperatures reduce the current response to applied magnetic fields due to increased carrier recombination and thermal excitations that compete with the magnetoresistive effect. Despite these factors, the sensitivity reduction is modest (almost two orders of magnitudes), demonstrating that our SiC-based NZFMR sensor remains robust for high-temperature magnetic field detection, even though not specifically designed for magnetometry applications.

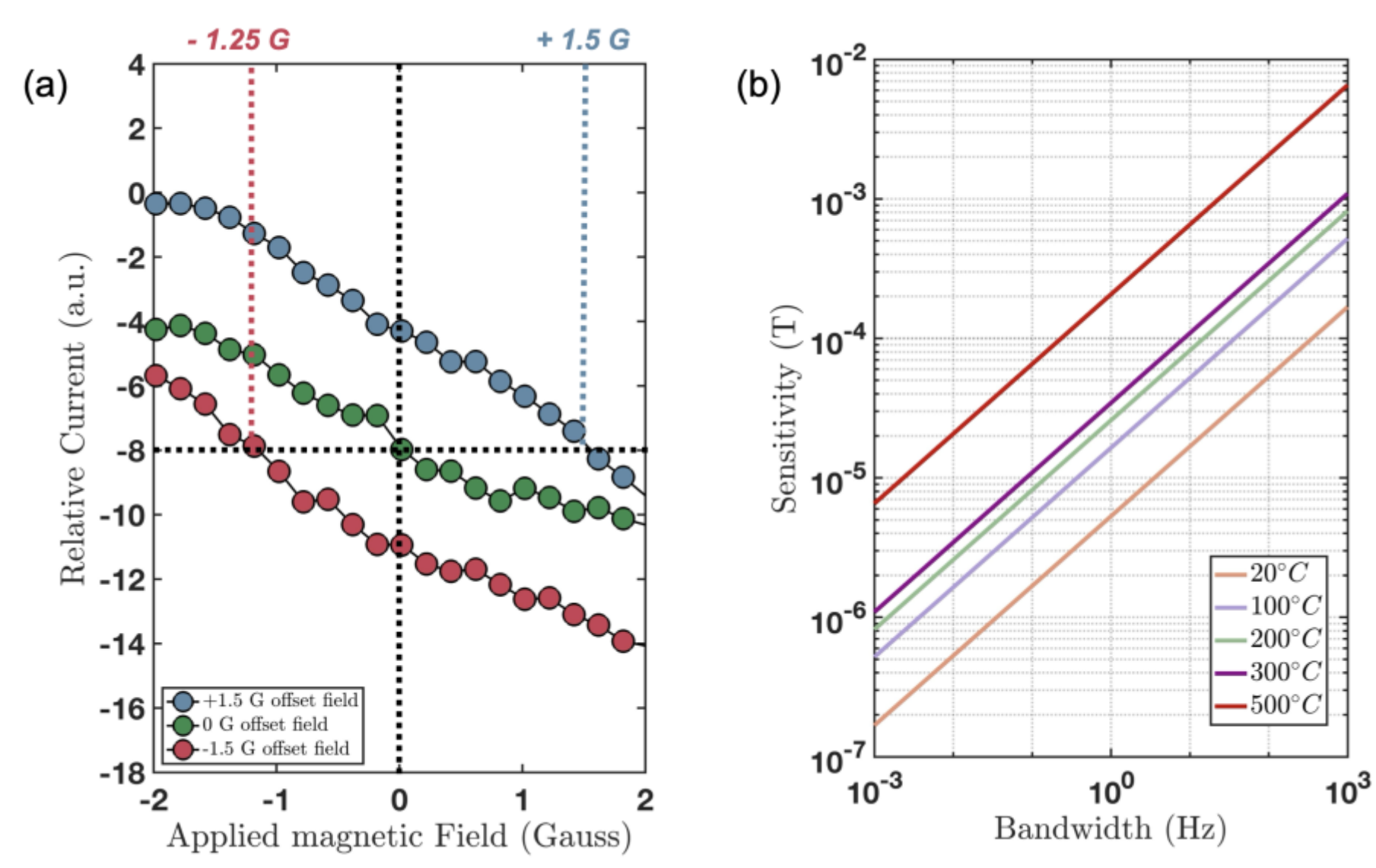

*Figure 4: (a) Demonstration of high-temperature relative magnetometry. Change in the device current with the magnetic field. Data are taken at 500 °C. Each curve has an average of twenty runs. (b) Magnetometry sensitivity estimations as a function of the bandwidth for different temperatures.*

## Conclusions

In conclusion, we demonstrate the feasibility of an all-electrical NZFMR-based relative magnetometry using SiC diodes for high-temperature applications. By integrating a commercially available SiC diode—originally not designed for magnetometry—onto a custom stack with high-temperature-ready modulation and nulling coils, we detect low magnetic fields at temperatures up to 500°C. Our system is cost-effective and scalable, thanks to in-house assembly using readily available electronic components, eliminating the need for specialized research instrumentations. Moreover, the compact form factor—approximately the size of a quarter—and low power consumption of under 0.5 W at 500°C for magnetic fields below 5 Gauss make this solution especially attractive. Its robust and low-SWaP design positions it as a promising option for all-electrical relative magnetometry in harsh environments.

Several strategies could be used to further improve this system's sensitivity, especially at high temperatures. First, designing SiC diodes specifically for magnetometry by engineering defect states and optimizing doping profiles could enhance the current response ΔI, reducing spin-independent recombination losses at elevated temperatures and, thus, a higher sensitivity. Additionally, using isotopically purified SiC material will minimize hyperfine interactions, leading to a narrower line width σ and an enhanced signal. Furthermore, advanced signal processing techniques, including adaptive filtering and noise reduction algorithms, could extract more accurate signals from noisy environments, improving the overall performance. By implementing these strategies, SiC-based NZFMR magnetometry can achieve superior sensitivity and reliability, positioning it as a leading solution to high-performance magnetic field sensing in extreme environments and a reliable technology for space exploration, automotive systems, and industrial applications.

## Acknowledgment

The room-temperature data are based on work supported by the National Aeronautics and Space Administration under Contract No. 80NSSC23CA145 issued through the SBIR/STTR Program. The high-temperature data are based on work supported by the Defense Advanced Research Projects Agency (DARPA) under Agreement No. HR0011-23-9-0114. We acknowledge useful discussions with M. E. Flatté, C. J. Cochrane, and H. Kraus.